\documentstyle[epsf,mnras_cite,times]{mn}
\input{epsf}
\begin{document}
\newcommand{\ba}{\begin{eqnarray}}
\newcommand{\ea}{\end{eqnarray}}
\newcommand{\be}{\begin{equation}}
\newcommand{\ee}{\end{equation}}
\newcommand{\nn}{\nonumber \\}
\newcommand{\vk}{{\bf{k}}}
\newcommand{\vkone}{{\bf{k}_1}}
\newcommand{\vktwo}{{\bf{k}_2}}
\newcommand{\vkthree}{{\bf{k}_3}}
\newcommand{\kvone}{{\bf{k}_1}}
\newcommand{\kvtwo}{{\bf{k}_2}}
\newcommand{\vr}{{\bf{r}}}
\newcommand{\vx}{{\bf{x}}}
\newcommand{\ls}{\mathrel{\raise1.16pt\hbox{$<$}\kern-7.0pt 
\lower3.06pt\hbox{{$\scriptstyle \sim$}}}}         
\newcommand{\gs}{\mathrel{\raise1.16pt\hbox{$>$}\kern-7.0pt 
\lower3.06pt\hbox{{$\scriptstyle \sim$}}}}         
\def\VEV#1{{\langle #1 \rangle}}
\long\def\comment#1{}
\def\hatn{{\bf \hat n}}
\def\fun#1#2{\lower3.6pt\vbox{\baselineskip0pt\lineskip.9pt
  \ialign{$\mathsurround=0pt#1\hfil##\hfil$\crcr#2\crcr\sim\crcr}}}
\def\lap{\mathrel{\mathpalette\fun <}}
\def\gap{\mathrel{\mathpalette\fun >}}
\def\Msun{{M_\odot}}
\def\Rvir{{R_{\rm vir}}}

\title{On Galaxy-Cluster Sizes and Temperatures}
\author[L. Verde, M. Kamionkowski, J. Mohr \& A. J. Benson]
{Licia Verde$^{1,2}$, Marc Kamionkowski$^{3}$, Joseph J. Mohr$^{4,5}$
\& Andrew J. Benson$^{3,6}$\\
$^1$ Institute for Astronomy, University of Edinburgh, Royal Observatory, 
Blackford Hill, Edinburgh EH9 3HJ, United Kingdom\\
$^{2}$ Dept. Astrophysical Sciences, Peyton Hall, Princeton University,
Princeton NJ 08544-1001, USA\\
$^{3}$ California Institute of Technology, Mail Code 130-33,
Pasadena, CA 91125, USA\\
$^{4}$ Department of Astronomy and Astrophysics, University of Chicago,
5640 South Ellis Avenue, Chicago, IL 60637-1433, USA\\
$^5$ Department of Astronomy, University of Illinois, 1002 W. Green, Urbana, IL
61801, USA\\
$^6$ Physics Department, University of Durham, Science Laboratories, South
Road, Durham DH1 3LE, United Kingdom\\}

%
\maketitle
\begin{abstract}
We show that the distribution of the sizes and temperatures of clusters can be
used to constrain cosmological models.  The size-temperature (ST) distribution
predicted in a flat Gaussian cluster-abundance-normalized $\Omega_0=0.3$ model
agrees well with the fairly tight ST relation observed.  A larger
power-spectrum amplitude $\sigma_8$ would give rise to a larger scatter about
the ST relation as would a larger value of $\Omega_0$ and/or long non-Gaussian
high-density tails in the probability density function.  For Gaussian initial
conditions, the ST distribution suggests a constraint $\sigma_8
\Omega_0^{0.26} \simeq 0.76$.  The ST relation is expected to get tighter at
high redshifts.  In the process, we derive a simple formula for the halo
formation-redshift distribution for non-Gaussian models.  We also suggest that
the discrepancy between the naive zero-redshift ST relation and that observed
may be due, at least in part, to the fact that lower-mass clusters form over a
wider range of redshifts.  An Appendix derives an equation for
the formation-redshift distribution of halos.
\end{abstract}
%
\section{Introduction}%

Galaxy clusters are now being widely used as probes of
cosmological and structure-formation models.  For example, the
abundance of galaxy clusters has been used to constrain the
amplitude $\sigma_8$ of the power spectrum and the
nonrelativistic-matter density $\Omega_0$ in models with an
initially Gaussian distribution of density perturbations
\cite{Evr89,HenArn91,BahCen92,BahCen93,Lil92,OukBla92,WhiEfsFre93,ViaLid96,Ekeetal96,ViaLid99}, 
as well as in models with long non-Gaussian tails
(\pcite{RobGawSil00}; hereafter RGS).

In this paper, we show that the scatter in galaxy cluster
scaling relations can be used to constrain cosmological and structure
formation models.  Specifically, we focus on the relatively small scatter
of the relation between X-ray isophotal size and
emission-weighted intracluster-medium mean
temperature $T_X$ demonstrated in Mohr \& Evrard (1997; hereafter ME97). 
We illustrate how this scatter should depend on
$\sigma_8$ and $\Omega_0$, and how it is affected by
the introduction of a non-Gaussian distribution of perturbations
with a long tail of high-density peaks.  Our work on the
size-temperature (ST) relation follows prior analytic work by \scite{KitSut96}
(although they focussed primarily on other cluster properties)
and employs the framework for relating the ST relation to the
underlying dark matter properties as discussed in Mohr et al. (2000; hereafter
M00).

\begin{figure}
\begin{center}
\setlength{\unitlength}{1mm}
\begin{picture}(90,55)
\includegraphics{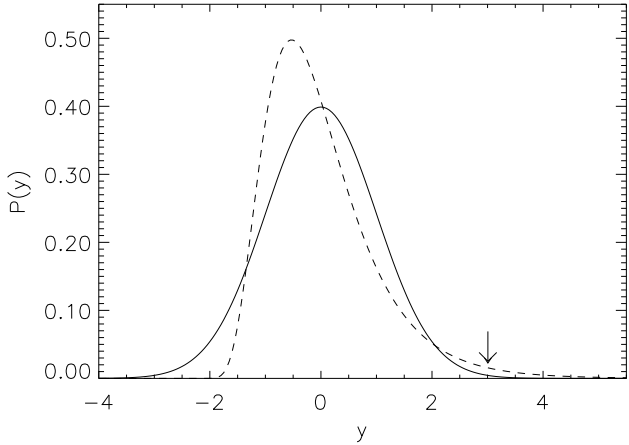}
\end{picture}
\end{center}
\caption{The solid curve shows a Gaussian distribution $P(y)$
     with unit variance, while the broken curve shows a
     non-Gaussian distribution with the same variance but 10
     times as many peaks with $y>3$.  This illustrates (a) how
     the cluster abundance can be dramatically enhanced with
     long non-Gaussian tails (since clusters form from rare
     peaks); and (b) that the dispersion of $y$ for $y>3$ is
     much larger for the non-Gaussian distribution than it is
     for the Gaussian distribution, and this will lead to a
     larger scatter in the formation redshifts and sizes of
     clusters of a given mass.}
\label{fig:distributions}
\end{figure}

The small scatter is heuristically expected if clusters form at
rare high-density peaks in a Gaussian primordial distribution.
Clusters that form earlier should be denser when they are first
virialized and so they should have smaller radii for a given mass, or
similarly, smaller radii for a given temperature.  In this way, any dispersion
in the formation redshifts for clusters of a given mass should yield a spread
in the ST relation.  If clusters come from rare Gaussian peaks,
then the spread in formation redshifts
should be small; given the rapidly dying Gaussian tails, it is unlikely that
any cluster of a given mass observed today was formed at a redshift much
earlier than the others.  However, if the distribution had long non-Gaussian
tails (as would be required to significantly boost the cluster
abundance) or if clusters formed from peaks
that were not quite so rare (e.g., $>2\sigma$ rather than
$>3\sigma$ peaks), then clusters of a given mass observed today
should have had a much broader distribution of formation
redshifts (see Fig. \ref{fig:distributions}) and thus a
much broader distribution of sizes (for a given mass or
temperature). 

We quantify these arguments using a spherical-top-hat-collapse model
to relate the virial radius and temperature of a cluster to its mass
and formation redshift.  We use the formation-redshift distribution for
Gaussian perturbations from \scite{sasaki}, and we generalize it for an
arbitrary initial density distribution (the derivation is
presented in an Appendix).  We use a Monte Carlo approach to
simulate the ST relation for a variety of parameters, and illustrate in
particular how it depends on $\sigma_8$, $\Omega_0$, and $G$, the non-Gaussian
multiplicative excess of $>3\sigma$ peaks introduced by RGS.  Our main results
are (a) the predicted scatter in the ST relation for Gaussian initial
conditions and favored cosmological parameters is found to be fairly
consistent with that observed; (b) $G\gap 5$
greatly overpredicts the scatter; (c) the scatter for the non-Gaussian initial
conditions required to make the cluster abundance consistent with an
Einstein-de-Sitter Universe (EdS) is also much larger than that observed.
Joint constraints from the cluster abundance and the ST relation on
$\sigma_8$, $G$, and $\Omega_0$ are discussed.  We show how the ST
relation should be altered for clusters at intermediate and high
redshifts. In the process we show that, because lower-mass clusters form over a
larger range of redshifts than higher-mass clusters, the expected ST relation
is steeper (and therefore more consistent with the observed relation) than the
naive expectation detailed in M00.  In the final Section, we
make some brief connections to the X-ray mass-temperature
relation and to the redshift evolution of the cluster abundance.

\section{Ingredients}
\subsection{Spherical-Collapse Model}
\label{sec:sthc}

\begin{figure}
\begin{center}
\setlength{\unitlength}{1mm}
\begin{picture}(90,55)
\includegraphics{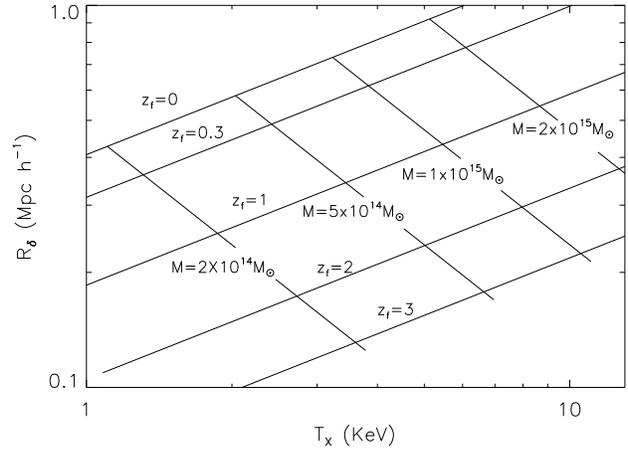}
\end{picture}
\end{center}
\caption{Mass and formation-redshift contours in the
size-temperature plane for $\Omega_0=0.3$ and $h=0.65$ obtained
{}from the spherical-top-hat model of gravitational collapse
discussed in the text.  It is clear from the figure that a
narrow (broad) spread in the formation redshift will yield a tight
(broad) ST relation.  For larger $\Omega_0$, the $z_f=0$ contour 
remains the same, but the spacing between equi-$z_f$ contours
increases.}
\label{fig:ST}
\end{figure}

We use the relations of \scite{KitSut96} to relate the cluster virial
radius  and virial temperature at formation time, $R_{\rm vir}$ and $T$, to the mass $M$ and formation
redshift $z_f$ (defined to be the redshift at which the cluster
collapses).
Fig. \ref{fig:ST} shows how this model assigns masses and
formation redshifts to clusters of given temperatures and sizes assuming that
$R_{\rm vir}\propto R_{\delta}$.

It is possible to connect more rigorously these cluster dark-matter properties
with the observable intracluster medium (ICM) properties in a manner similar
to that outlined in M00.  Specifically, we assume that $T_X$ is the virial
temperature (e.g. Evrard, Metzler \& Navarro 1996; Frenk et al. 2000; Bower et
al. 2000).  We transform from the virial radius at $z_f$ to the X-ray isophotal size
$R$ using $ R\propto R_{\rm vir}^{4/3} f_{ICM}^{2/3},$ where
$f_{ICM}$ is the ICM
mass fraction (M00 eqns. (8) and (10)).  The dependence on $f_{ICM}$
should be included because variations in $f_{ICM}$ with mass
are observed (e.g. Mohr, Mathiesen \& Evrard 1999; David, Jones \& Forman
1995)  and could alter the slope of the ST relation.  In the following
analysis, we assume $f_{ICM}\propto T_X^{0.34}$; however the results of
Figs. 4 and 5 are no more than {\it very weakly} dependent on
the $f_{ICM}$ functional form.

We normalize the simulated ST relation to the observations by
fixing the constant of proportionality so that no observed cluster in the
local sample lies above the $z_f=0$ line (see Figs 2 and 3).

\subsection{Distribution of Halo Masses}

Numerical simulations tell us that the Press-Schechter (PS)
approach \cite{PreSch74} provides a
reasonable approximation  for the abundance of cluster size halos of a given
mass at any given epoch for Gaussian initial
conditions (e.g. Lacey \& Cole 1994, Gross et al. 1998, Lee \& Shandarin 1999), and for a few non-Gaussian initial
conditions that have been explored with simulations \cite{RobBak99}.
In the PS approach the number per comoving volume of halos with masses
between $M$ and $M+dM$ at redshift $z$ is (e.g., Lucchin \& Matarrese 1988; RGS),
\be
     {dn \over dM} \, dM = {f \rho_b \over M}
     P\Bigl(y(M,z)\Bigr) {\partial y(M,z) \over \partial M}\,
     dM,
\label{eq:PS}
\ee
where $\rho_b$ is the background density, $P(y)$ is the
primordial probability distribution function normalized to unit
variance.  The argument $y=\delta(z)/\sigma_M$, and
$\delta(z)=\delta(z)/D(z)$ where $\delta_c(z)$ is the critical
overdensity for collapse (see \pcite{KitSut96} for accurate
analytic fits), and $D(z)$ is the linear-theory growth factor.
Here, $\sigma_M$ is the current root-variance of spheres that
enclose an average mass $M$, and $f=\int_0^\infty\,P(y)\, dy$.

\begin{figure*}
\begin{center}
\setlength{\unitlength}{1mm}
\begin{picture}(90,55)
\includegraphics{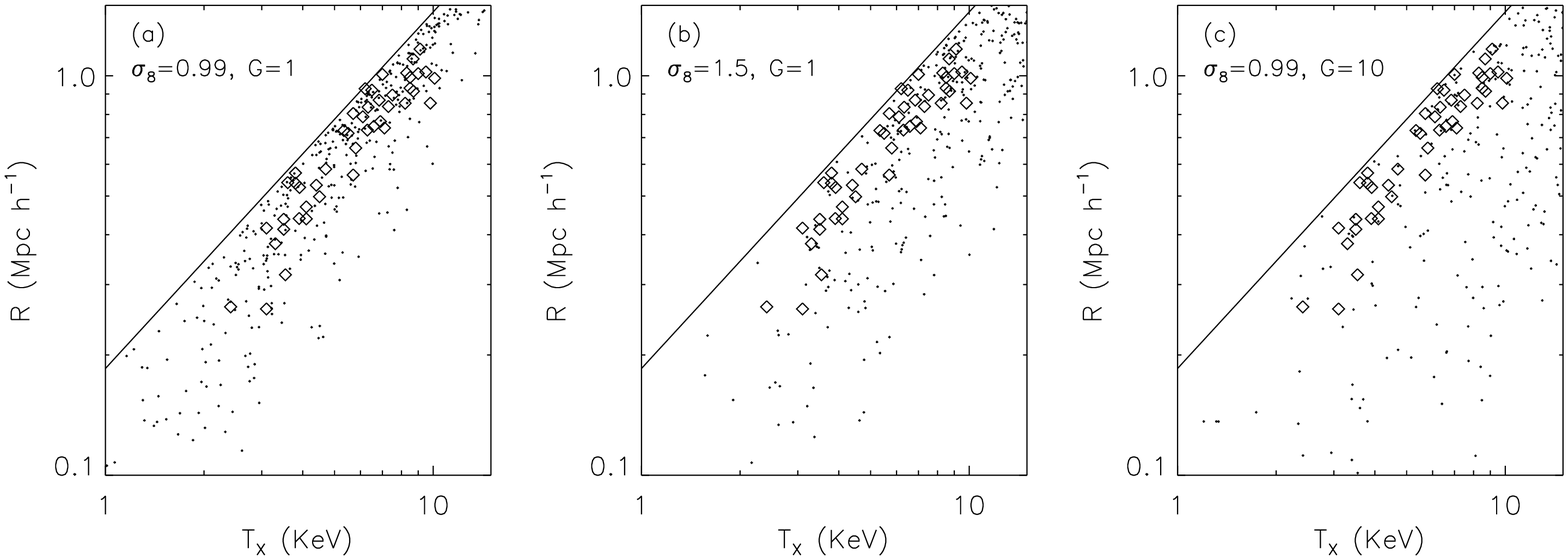}
\end{picture}
\end{center}
\caption{(a) ST distribution for LCDM and
     $\sigma_8=0.99$ and Gaussian initial conditions.  Each dot
     represents a simulated cluster, while the diamonds are
     data from M00.  The line shows the ST
     relation expected for clusters that form today, at redshift 
     $z=0$.  (b) shows the same
     except that here we use $\sigma_8=1.5$.  (c) shows the same
     as in (a) but with the non-Gaussian distribution of
     RGS with $G=10$.} 
\label{fig:basicresults}
\end{figure*}
\subsection{Distribution of Formation Redshifts}

The objects of mass $M$ observed at some given redshift $z_o$ underwent
collapse at a variety of formation redshifts $z_f >z_o$.  
Sasaki (1994) has shown how the PS formalism leads to an expression for the
formation-redshift distribution under the assumption of Gaussian initial
conditions and that the merger rate has no characteristic mass scale. 
His derivation can be generalized in a straightforward
fashion to arbitrary $P(y)$.  Doing so (see Appendix), we find
the distribution (normalized to unity) of formation redshifts
$z_f$ for halos of mass $M$ observed at redshift $z_o$ to be,
\begin{eqnarray}
     {df \over dz_f} = P'\bigl(y(M,z_f)\bigr) 
     {\partial y(M,z_f) \over \partial
     z_f}\left[P\Bigl(y(M,z_0)\Bigr)\right]^{-1}
\label{eq:zdist}
\end{eqnarray}
where $P'(y) \equiv dP/dy$.
Lacey \& Cole (1993,1994) have presented an alternative, but
somewhat more complicated, formation-redshift distribution that
improves upon Sasaki's assumption of self-similar merging.
We will leave the implementation of this alternative
distribution and a discussion of the formalism introduced by
\scite{Percivaletal}, to future work, but note that our
preliminary investigations, as well as previous results
\cite{ViaLid96,Buc00}, indicate that the predictions of these 
models do not differ considerably for cluster-mass halos. 

\subsection{Preliminary Estimates}

It is straightforward to roughly estimate the effects of
non-Gaussian tails on the ST-relation scatter.
For a rapidly dying distribution
$P(y)$, the controlling factor in ${dn/dz_f}$ will be $P'(y)$.
For a Gaussian $P(y)$, the root-variance of $y$ is 0.282 for the
distribution $P'(y)$ for values of $y>3$, and the mean value of
$y$ is 3.30.  For an EdS model,
$y=1.69(1+z_f)/\sigma_M$, and $(1+z)^{-1} \propto R_{\rm vir}$.
Thus, $\sigma_R/R = (4/3) (\sigma_{R_{\rm vir}}/R_{\rm vir})
\simeq (4/3) (\sigma_y/y)=0.113$ for a Gaussian distribution, in 
surprisingly good agreement with the estimate of the intrinsic scatter of
10\% in the ST relation (ME97).  For the RGS
distribution with $G=10$, the root-variance is $0.896$ and the
mean value of $y$ is 3.87 leading to $\sigma_{R_{\rm vir}}/R_{\rm vir}\simeq 0.31$,
more than twice the observed scatter.  Below we will quantify
this far more precisely.

\section{Results}

For any given $\Omega_0$, $\sigma_8$, and $G$, we perform a 
Monte Carlo realization of 400 clusters with the
mass and formation-redshift distributions given above.  We then
assign to each of these clusters a size and temperature
as outlined in Section \ref{sec:sthc}.  The ME97
sample to which we compare our calculations is a flux-limited
sample. Within this sample, the probability of finding a cluster of luminosity
$L_X$ goes as $L_X^{1.5}$, and $L_X$ is observed to go as
roughly $T^{2.5}$ to $T^3$ \cite{Davetal93,ArnEvr99}, so the flux
limit is essentially a virial-temperature weighting of
$T^{3.75}$ to $T^{4.5}$.  We thus subject our simulated
population of clusters to a $T^{3.75}$ weighting; our results 
are not significantly altered for the steeper weighting $T^{4.5}$.

Fig. \ref{fig:basicresults}(a) shows the results of our Monte Carlo 
for a flat $\Omega_0=0.3$ model (LCDM) with  the value
$\sigma_8=0.99$ inferred from
the cluster abundance \cite{ViaLid99} and a Gaussian
distribution.  The data points from ME97 are overlaid.  We used a Hubble
parameter $h=0.65$, but the results are essentially unaltered
for different plausible values of $h$.
Fig. \ref{fig:basicresults}(b) illustrates that the scatter in
the ST relation is increased if the power-spectrum
normalization is higher.  In this case, clusters are not quite
as rare, and they form over a larger range of
redshifts. Fig. \ref{fig:basicresults}(c) shows how the scatter
is increased as the abundance of high-density peaks is
increased.  In this case, clusters observed today are also formed
over a broader range of redshifts.  At this point, we note the
apparent similarity between
the predictions of the ST distribution of the
cluster-abundance-normalized Gaussian LCDM model and the
data;  the scatter about the ST
relation would be broadened considerably with a higher
$\sigma_8$ or with a highly non-Gaussian model.

\begin{figure}
\begin{center}
\setlength{\unitlength}{1mm}
\begin{picture}(90,55)
\includegraphics{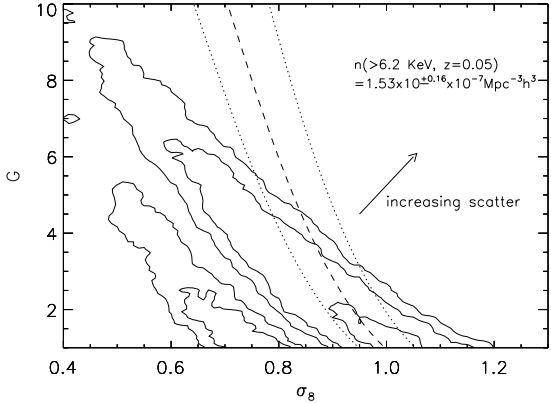}
\end{picture}
\end{center}
\caption{The heavy solid curves show confidence levels
     suggested by the ST data in the
     $\sigma_8$-$G$ parameter space for the LCDM model, and the
     light curves show the same for the EdS
     model.  The dashed curve shows the contour suggested by the
     central value [$n(>6.2\,{\rm
     keV},z=0.05)=1.53\times 10^{\pm 0.16}\times10^{-7}\,{\rm
     Mpc}^{-3}\,h^3$] of the local cluster abundance for
     $\Omega_0=0.3$, while the dotted curves indicate contours
     for the upper and lower observational limits to the cluster
     abundance.} 
\label{fig:likelihood}
\end{figure}

To make these arguments more quantitative as well as survey a
larger range of parameters, we have simulated
ST relations for a variety of models in the $\sigma_8$-$G$
parameter space for both EdS and LCDM models and
then used a 2D Kolmogorov-Smirnov (KS) test
\cite{Pea83,Preetal96} to compare these with the data.
Fig. \ref{fig:likelihood} shows the resulting contours of
constant KS significance levels for both 
$\Omega_0=0.3$ and $\Omega_0=1$.  The
results suggest that the Gaussian cluster-abundance-normalized
($\sigma_8=0.99$) LCDM model provides a good fit to the data.

We heuristically expect that the dependence of the ST scatter on 
cosmological parameters/models should be similar to that of the
cluster abundance; if the peaks that give rise to clusters are
rare, we expect little scatter and {\it vice versa} if clusters
are more common.  The contours of fixed cluster abundance in
Fig. \ref{fig:likelihood} indicate that this is qualitatively
correct.  We obtain these
curves by using a cluster abundance $n(>6.2\, {\rm keV},
z=0.05)=1.53\times 10^{\pm 0.16}\times10^{-7}\,{\rm Mpc}^{-3}\,h^3$ 
\cite{ViaLid99} and integrating eqn. (\ref{eq:PS}) up from
the mass associated with a temperature 6.3~keV and a formation
redshift $z_f=0$. However, the detailed results also seem to indicate that
if $\Omega_0$ is fixed, the ST distribution and cluster
abundance can be used in tandem to break the degeneracy between
$G$ and $\sigma_8$.  In fact, combining the two constraints
already seems to rule out large deviations from Gaussianity.

\begin{figure}
\begin{center}
\setlength{\unitlength}{1mm}
\begin{picture}(90,55)
\includegraphics{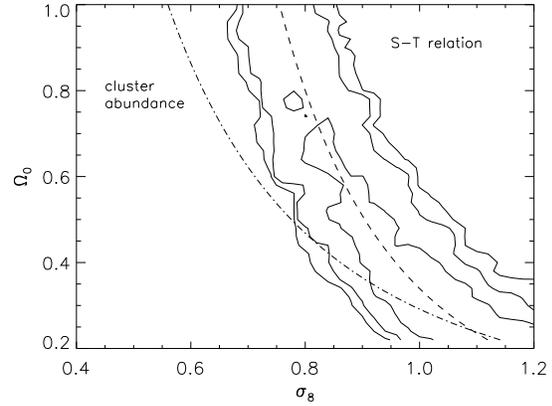}
\end{picture}
\end{center}
\caption{The heavy solid curves show likelihood contours
     suggested by the ST data for Gaussian initial conditions in the
     $\sigma_8$-$\Omega_0$ parameter space.
     The dot-dashed curve shows the contour preferred by the local
     cluster abundance as suggested by \protect\scite{ViaLid99}, while the dashed curve shows the fit to our
     ST constraint.}
\label{fig:SOlikelihood}
\end{figure}

Fig. \ref{fig:SOlikelihood} shows the regions of
$\sigma_8$-$\Omega_0$ parameter space preferred by the ST
relation, as well as the curve in this parameter space
suggested by the cluster abundance.  For fixed $\sigma_8$, the
ST scatter increases as $\Omega_0$ increases.  At first, this
might seem discrepant with the well-known result that the range
of formation redshifts is narrower for larger $\Omega_0$ for
cluster-abundance-normalized models.  However, this narrowing of 
the formation-redshift distribution with increasing $\Omega_0$
is not quite as dramatic if we fix $\sigma_8$ instead of the
cluster abundance.  More importantly, the
spherical-top-hat-collapse dynamics leads to a broader spacing
between the equi-$z_f$ contours in Fig. \ref{fig:ST}, and this
is responsible for increasing the ST scatter as $\Omega_0$ is
increased with fixed $\sigma_8$; in other words, the relationship between $R$
and $T$ evolves more rapidly with redshift in higher $\Omega_0$ models.  

From the results in Fig. \ref{fig:SOlikelihood}, we can
approximate an ST constraint, $\sigma_8=0.76\,\Omega_0^{-0.26}$,
as compared with the cluster-abundance constraint,
$\sigma_8=0.56 \, \Omega_0^{-0.47}$ \cite{ViaLid99}.  
The region of overlap between the cluster-abundance constraint
and the ST relation lies at low values of $\Omega_0$, low values
of non-Gaussianity, and slightly higher values of $\sigma_8$.
\vspace*{-0.35cm}
\subsection{An Einstein-de-Sitter Universe?}
\begin{figure}
\begin{center}
\setlength{\unitlength}{1mm}
\begin{picture}(90,55)
\includegraphics{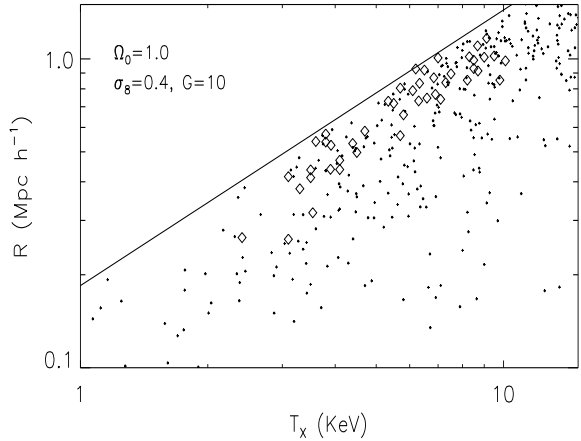}
\end{picture}
\end{center}
\caption{The ST distribution for $\Omega_0=1$ with
     $\sigma_8=0.4$ and $G=10$, one of the combinations of
     parameter values that yield the correct cluster abundance
     for an EdS Universe.  The predicted scatter in the
     ST relation is considerably larger than that
     observed.}
\label{fig:EdS}
\end{figure}

RGS were able to identify  for an EdS model, a region in the
$\sigma_8$-$G$ parameter space near $\sigma_8=0.4$ and $G=10$ in 
which the predicted cluster abundance was found to agree with
that observed.  Fig. \ref{fig:EdS} shows that these parameter
choices predict far too much scatter in the ST
relation. Allowing for additional sources of scatter
in this simulated ST relation would only increase the discrepancy between
the model and the observations.

\subsection{High and Intermediate Redshift Results}

\begin{figure}
\begin{center}
\setlength{\unitlength}{1mm}
\begin{picture}(90,80)
\includegraphics{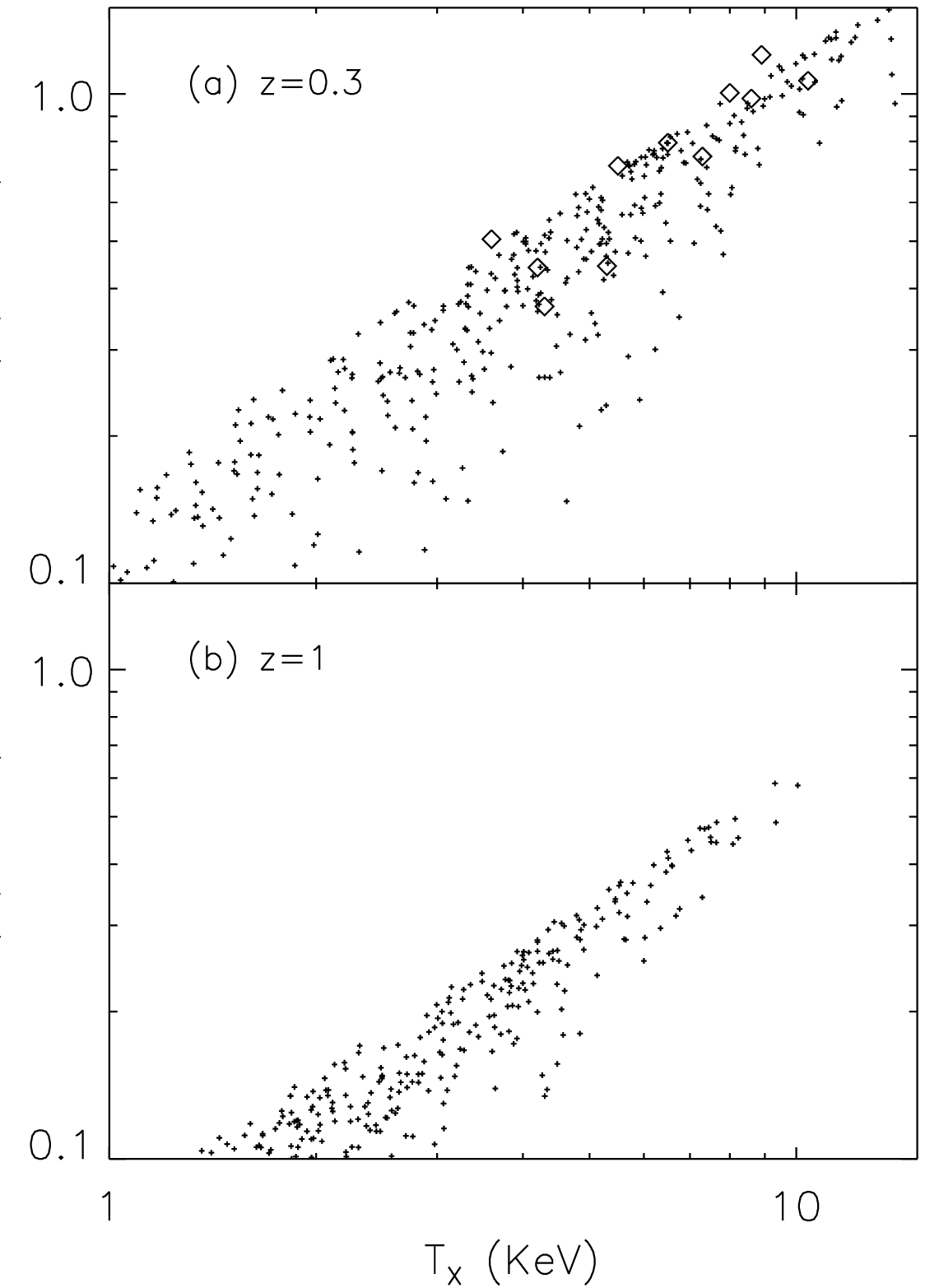}
\end{picture}
\end{center}
\caption{The ST relation for the LCDM model with $\sigma_8=0.99$ and Gaussian
     perturbations  for (a) clusters at $z=0.3$ and (b) $z=1$.  This
     illustrates how the scatter should decrease with redshift. 
The data from \protect\scite{Mohetal00} (that have median redshift 0.3,
but cover the range $0.19 <z <0.54$) are shown in (a).}
\label{fig:highz}
\end{figure}

Clusters that exist at higher redshifts must form from even
higher-density peaks than those today.  Thus, in a Gaussian
model, the scatter in their formation redshifts and thus in
their sizes should be even smaller.  This is illustrated in
Fig. \ref{fig:highz}.  The canonical-model predictions shown in
Fig. \ref{fig:highz}(a) for $z\simeq 0.3$ seem to be in
relatively good agreement with the cluster sample observed so
far.  Fig. \ref{fig:highz}(b) shows that the scatter in the
ST relation for the canonical model should be {\it 
very} small.  Even though the sample of such high-redshift
clusters is expected to be small, the predicted scatter is so
small that measurement of the sizes of only a handful of
clusters could put strong constraints on different sources of scatter
(e.g., non-Gaussianity, measurement uncertainties, mergers,
galaxy feedback, etc.).

\subsection{A size-temperature anomaly?}

The ST relation of the low-redshift X-ray flux limited cluster sample has a
slope of $m\sim1$, which is considerably steeper than the $m=2/3$ slope expected in a
model where all clusters are assumed to have formed at the redshift of
observation (ME97, M00).  ME97 suggest that a possible explanation
for this steeper than expected scaling relation is provided by galaxy
feedback.  Fig. \ref{fig:basicresults} illustrates that the discrepancy may be
due only, or at least in part, to the fact that lower-mass clusters form over
a broader range of redshifts, and thus will in general have smaller sizes than
they would if they all formed very recently.  Visual inspection of
Fig. \ref{fig:basicresults} suggests that this is a plausible explanation,
especially when the small-number statistics of the observational sample are
taken into account.  Moreover, the relatively strong dependence of the ST
scatter on $\sigma_8$ indicate that better agreement than shown in
Fig. \ref{fig:basicresults} could be obtained with a slightly different value
of $\sigma_8$ and/or $\Omega_0$ (cf., Fig. \ref{fig:SOlikelihood}).  The
apparent disagreement with ME97's feedback-free numerical simulations, which
show an ST scaling close to the naive scaling (but still steeper in 3 of the 4
cosmologies tested), may have been due to this $\sigma_8$ and $\Omega_0$
dependence and/or the relatively small-number statistics of their simulations
sample.  Thus, the apparent deviation of the ST relation slope from the
$m=2/3$ expectation  is not quite so anomalous.

\vspace*{-0.3cm}
\section{Discussion}
We have calculated the ST distribution of clusters 
with a simple analytic model and focussed in particular on the
dependence on the power-spectrum amplitude ($\sigma_8$) and the degree of
non-Gaussianity ($G$).  We find a fairly sensitive dependence of the
ST relation scatter on these two parameters.  Thus, the tightness of the
ST relation can be used to place valuable
constraints on these parameters, as well as on other
cosmological parameters.  The canonical
cluster-abundance-normalized $\Omega_0=0.3$ model predicts an
ST relation consistent with that observed, but 
a $\sigma_8$ much larger or smaller would be
inconsistent as would a non-Gaussian model that predicts a
significant excess of $>3\sigma$ high-density 
peaks.\footnote{However, small deviations from Gaussianity have huge impact on
clusters at $z_o\gg 0$ (e.g., Matarrese, Verde \& Jimenez 2000); for example the small non-Gaussinanity
(e.g. $G \simeq 2$) required to accommodate the existence of the MS1054-03
cluster within the LCDM model (Verde et al. 2000b) is consistent 
with the ST constraint.}
The constraints to $\sigma_8$, $G$, and $\Omega_0$ that arise from
the ST distribution should be qualitatively similar to
those from cluster abundances, but our preliminary calculations
suggest that they may be sufficiently different to provide
complementary constraints.  The ST relation should become
increasingly tight at larger redshifts.  Our results also
suggest that the discrepancy between the naive $z=0$
ST relation and the observed ST
relation may be due, at least in part, to the fact that
lower-mass clusters observed today have formed over a larger
range of redshifts than higher mass clusters.

The fact that lower-mass clusters tend to form over a broader redshift range
than higher-mass clusters will also tend to steepen the $M_{\rm vir}$-$T$
relation beyond the self-similar expectation of $m=2/3$.  Numerical
simulations of structure formation within models with non-Gaussian initial
conditions or low-$\Omega_0$ open models ought to exhibit this effect.  The
OCDM256 portion of Fig 4. in  Bryan \& Norman (1998) indicates that low-mass
clusters fall systematically below the best fit $M_{\rm vir}$-$T$ relation,
consistent with our expectation. It should be emphasized that, in this
particular study, the low-mass systems are
composed of far fewer particles than the high-mass systems, providing
another plausible explanation for structural differences.  Further work to
investigate departures from self similarity in the cluster population which
naturally arise from the spread in formation epochs is clearly
required.  

Since the overdensity-peak amplitude at which a cluster can form
increases at higher redshift, the redshift evolution of the
cluster abundance depends on the shape of the primordial density
distribution function at high peaks just as the ST scatter
does.  Thus, if $\Omega_0$ is fixed, it should be possible to
reconstruct the cluster-abundance evolution from the scatter in
the ST relation for local clusters.

Although we have used cluster sizes inferred from X-rays to compare with
theoretical calculations, the same could be done for the sizes of clusters
measured via the Sunyaev-Zeldovich effect, either with or without redshift
information \cite{Kametal00}.  Of course, there will invariably be some
cluster-formation physics that our current analysis has left out, and
numerical simulations may have an advantage in this regard.  Note that the
only source of scatter in our simulated ST relations is the range of formation
epoch, whereas other sources of stochasticity (e.g., measurement uncertainties,
mergers, galaxy feedback, etc.) might increase the scatter.  However, with our
analytic approach, we are able to rule out models that overpredict the scatter;
we can sift far more rapidly through a variety of cosmological models and
parameters, study the dependence of the ST distributions on these models and
parameters, and gain some intuitive feel for how the results arise.  By doing
so, we hope to have established that cluster sizes can provide a valuable new
probe of cosmological models.

\section*{ACKNOWLEDGMENTS}
We thank P. Goldreich for asking questions that stimulated this
investigation.  This research was supported in part by the NSF
under grant no. PHY94-07194 to the Institute for Theoretical
Physics (Santa Barbara), where part of this research was completed.
LV was supported by a TMR grant and thanks Caltech for hospitality. MK was
supported in part by NSF AST-0096023, NASA NAG5-8506, and DoE
DE-FG03-92-ER40701.  JJM is supported by Chandra Fellowship grant PF8-1003,
awarded through the Chandra Science Center.  The Chandra Science Center is
operated by the Smithsonian Astrophysical Observatory for NASA under contract
NAS8-39073. AJB was supported by a PPARC studentship. 

\section*{APPENDIX}

Taking the derivative of eqn.~(\ref{eq:PS}) with respect to
redshift, we obtain (hereafter we do not explicitly show the
$(M,z)$ dependencies where they are obvious)
\begin{eqnarray}
{{\rm d}^2n \over {\rm d}M{\rm d}z}&=&\frac{f \rho_b}{M}
\left[\frac{\partial P}{\partial y}\frac{\partial
y}{\partial z}\frac{\partial y}{\partial M} + P(y)\frac{\partial }{\partial
z}\frac{\partial y}{\partial M}\right] \label{eq:PSdiff} \\
 &=&-\frac{f \rho_b}{M} \frac{1}{\sigma^2}\frac{\partial
\delta}{\partial z} \frac{\partial \sigma}{\partial M}\left[\frac{\partial
P}{\partial y}\frac{\delta}{\sigma} + P(y)\right] \label{eq:PSdiff1} \\
       &=& {{\rm d}^2n_{\rm form}
\over {\rm d}M{\rm d}z} - {{\rm d}^2n_{\rm dest} \over {\rm
d}M{\rm d}z}.
\label{eq:totrate}
\end{eqnarray}
In the last line we have equated the total rate of change to the 
difference between a formation rate and a destruction rate  (the
latter being due to objects merging to form larger objects).
These can be expressed as
\begin{equation}
{{\rm d}^2n_{\rm form} \over {\rm d}M{\rm d}z}=\int_{M_{\rm min}}^M {{\rm d} n
\over {\rm d}M} Q(M,M^\prime;z){\rm d}M^\prime ,
\label{eq:formrate}
\end{equation}
where $Q(M,M^\prime;z)$ is the probability that an object of mass
$M^\prime$ is one of the merging components when an object of mass $M$
forms and $M_{\rm min}$ is introduced to prevent the integral from
diverging, and ${\rm d}^2n_{\rm dest} / {\rm d}M{\rm
d}z=\phi(M,z) ({\rm d} n /{\rm d}M)$,
where the function $\phi(M,z)$ can be interpreted as the destruction
rate per bound object.
Sasaki assumes that $\phi(M,z)$ can be expressed as
$\phi(M,z)=M^\alpha \tilde{\phi}(z)$ (implying that the destruction
rate has no characteristic mass scale). Using
eqn.~(\ref{eq:totrate}), we can write
\begin{equation}
\tilde{\phi}(z) = {-{\rm d}^2n/{\rm d}M{\rm d}z + {\rm d}^2n_{\rm form}/{\rm d}M{\rm d}z \over {\rm d}n/{\rm d}M M^\alpha}.
\end{equation}
Since the left-hand side of this equation depends only upon $z$ the
right-hand side must be independent of $M$ and so may be evaluated at
a very small mass $M_{\rm min}$. Since the formation rate is
zero at $M_{\rm min}$ (see eqn.~\ref{eq:formrate}), this leaves
\begin{equation}
\tilde{\phi}(z) = - {{\rm d}^2n/{\rm d}M{\rm d}z(M_{\rm min},z) \over {\rm d}n/{\rm d}M(M_{\rm min},z) M_{\rm min}^\alpha}.
\end{equation}
Substituting eqns.~(\ref{eq:PS}) and (\ref{eq:PSdiff1}) into this
expression gives
\begin{equation}
\tilde{\phi}(z)=\frac{1}{\delta}\frac{\partial \delta}{\partial z}M_{\rm min}^{-\alpha}\left[
\frac{1}{P[y(M_{\rm min})]} \frac{\partial P[y(M_{\rm min})]}{\partial y} \frac{\delta}{\sigma(M_{\rm min})}+1\right]
\label{eq:tphi}
\end{equation}
For a hierarchical clustering model, $\lim_{M \to 0} \sigma^2(M)=\infty$,
so if we take the limit $M_{\rm min}\rightarrow 0$
eqn.~(\ref{eq:tphi}) will be $0$ or $\infty$ unless $\alpha =0$,
forcing the choice $\alpha=0$ upon us such that
$\tilde{\phi}(z)=(1/\delta)({\rm d}\delta/{\rm d}z)$. Substituting
this expression and eqn.~(\ref{eq:PSdiff}) into
eqn.~(\ref{eq:totrate}), we find that the formation rate is given by
\begin{equation}
{{\rm d}^2n_{\rm form} \over {\rm d}M{\rm d}z}=-\frac{f \rho_b}{M} \frac{1}{\sigma^2}\frac{\partial
\delta}{\partial z} \frac{\partial \sigma}{\partial M} \frac{\partial
P}{\partial y}\frac{\delta}{\sigma}.
\label{eq:formrate1}
\end{equation}
This is the rate of formation of bound objects of mass $M$ and
redshift $z$, but we wish to know what fraction of these objects will
survive until the redshift of observation. Using our definition of
$\phi(z)$ the number of objects of mass $M$ which formed at $z_{\rm f}$ must
evolve with redshift as ${\rm d}N/ {\rm d}z = \phi(M,z)N$ such
that the fraction remaining by $z_{\rm o}(<z_{\rm f})$ is $
f(z_{\rm f},z_{\rm o}) = \exp \int_{z_{\rm
f}}^{z_{\rm o}}\phi(z){\rm d}z = \delta(z_{\rm o})/ \delta(z_{\rm f})$.
The number of objects of mass $M$, which formed at redshift
$z_{\rm f}$ and which survive until redshift $z_{\rm o}$ is given by
the product of this expression and eqn.~(\ref{eq:formrate1});
i.e.,
\begin{equation}
{{\rm d}^2n \over {\rm d}M{\rm d}z}={f \rho_{\rm b} \over M}
{\delta(z_{\rm o})\over\delta(z_{\rm f})}{\partial y \over
\partial z}(z_{\rm f}) {\partial y \over \partial M}(z_{\rm f})
{\partial P \over \partial y}(z_{\rm f}).
\label{eq:formrate2}
\end{equation}
Noting that $(\partial y/\partial M)_{z_{\rm
f}}/(\partial y/\partial M)_{z_{\rm o}}=\delta(z_{\rm
f})/\delta(z_{\rm o})$, we obtain our final result,
eqn.~(\ref{eq:zdist}), by dividing eqn.~(\ref{eq:formrate2}) by
eqn.~(\ref{eq:PS}).

\end{document}